\documentclass[twocolumn]{autart}    

\usepackage{graphicx}          
\usepackage[dvips]{epsfig}    

\usepackage{graphicx}
\usepackage{epstopdf}
\usepackage{multicol}

\usepackage{amsfonts}

\usepackage{amsmath,amssymb,bm}    
\makeatletter
\renewcommand\normalsize{%
	\@setfontsize\normalsize\@xpt\@xiipt
	\abovedisplayskip 1\p@ \@plus1\p@ \@minus6\p@
	\abovedisplayshortskip \z@ \@plus3\p@
	\belowdisplayshortskip 3\p@ \@plus3\p@ \@minus3\p@
	\belowdisplayskip \abovedisplayskip
	\let\@listi\@listI}
\makeatother

\makeatletter
\renewcommand\small{%
	\@setfontsize\small{8pt}{10.5pt}
	\abovedisplayskip 1\p@ \@plus1\p@ \@minus6\p@
	\abovedisplayshortskip \z@ \@plus3\p@
	\belowdisplayshortskip 3\p@ \@plus3\p@ \@minus3\p@
	\belowdisplayskip \abovedisplayskip
	\let\@listi\@listI}
\makeatother

\makeatletter
\renewcommand\footnotesize{%
	\@setfontsize\footnotesize{7pt}{9pt}
	\abovedisplayskip 1\p@ \@plus1\p@ \@minus6\p@
	\abovedisplayshortskip \z@ \@plus3\p@
	\belowdisplayshortskip 3\p@ \@plus3\p@ \@minus3\p@
	\belowdisplayskip \abovedisplayskip
	\let\@listi\@listI}
\makeatother

\makeatletter
\renewcommand\scriptsize{%
	\@setfontsize\scriptsize{6pt}{7.875pt}
	\abovedisplayskip 1\p@ \@plus1\p@ \@minus6\p@
	\abovedisplayshortskip \z@ \@plus3\p@
	\belowdisplayshortskip 3\p@ \@plus3\p@ \@minus3\p@
	\belowdisplayskip \abovedisplayskip
	\let\@listi\@listI}
\makeatother

\makeatletter
\renewcommand\tiny{%
	\@setfontsize\tiny{4pt}{5.25pt}
	\abovedisplayskip 1\p@ \@plus1\p@ \@minus6\p@
	\abovedisplayshortskip \z@ \@plus3\p@
	\belowdisplayshortskip 3\p@ \@plus3\p@ \@minus3\p@
	\belowdisplayskip \abovedisplayskip
	\let\@listi\@listI}
\makeatother

\usepackage[toc,page]{appendix}
\usepackage{caption2}
\usepackage{enumerate}

\usepackage{algorithm}
\usepackage{algorithmic}
\usepackage{booktabs}
\usepackage{setspace}
\usepackage{stfloats}

\usepackage[colorlinks,
linkcolor=cyan,
anchorcolor=cyan,
citecolor=cyan,
hyperfigures= TRUE,
           ]{hyperref}

\usepackage{url}
\usepackage[round,sort]{natbib}
\def\QEDopen{{\setlength{\fboxsep}{0pt}\setlength{\fboxrule}{0.2pt}\fbox{\rule[0pt]{0pt}{1.3ex}\rule[0pt]{1.3ex}{0pt}}}} 
\graphicspath{{./}{figs/}{fig/}}

\usepackage{color}
\usepackage{xcolor}

\usepackage{appendix}

\newtheorem{define}{Definition}

\begin{document}

\begin{frontmatter}


\title{Event-triggered Observability: A Set-membership Perspective}


\author[BeiJing]{Kaikai~Zheng}\ead{kaikai.zheng@bit.edu.cn},    
\author[BeiJing]{Dawei~Shi}\textsuperscript{,\hspace{-0.2em}}\footnotemark[1]\thanks{Corresponding author.}\hspace{-0.25em}\ead{daweishi@bit.edu.cn},             
\author[Alberta]{Tongwen~Chen}\ead{tchen@ualberta.ca}  

\address[BeiJing]{State Key Laboratory of Intelligent Control and Decision of Complex Systems, School of Automation, Beijing Institute of Technology, Beijing, 100081, China}             
\address[Alberta]{Department of Electrical and Computer Engineering, University of Alberta, Edmonton, Alberta, Canada T6G 1H9}        

\begin{keyword}                           
Observability, Event-triggered state estimation, Set-membership filtering.              
\end{keyword}                             

\begin{abstract}   
This work attempts to discuss the observability of linear time-invariant systems with event-triggered measurements. A new notion of observability, namely, $\epsilon$-observability is defined with parameter $\epsilon$, which relates to the worst-case performance of inferring the initial state based on not only the received measurement but also the implicit information in the event-triggering conditions at no-event instants. A criterion is developed to test the proposed $\epsilon$-observability of discrete-time linear systems, based on which an iterative event-triggered set-membership observer is designed to evaluate a set containing all possible values of the state. The proposed set-membership observer is designed as the outer approximation of the ellipsoids predicted based on previous state estimates and the ellipsoids inferred by fusing the received measurement and communication conditions, which is optimal in the sense of trace at each step and is proved to be asymptotically bounded. The efficiency of the proposed event-triggered set-membership state observer is verified by numerical experiments.
\end{abstract}

\end{frontmatter}

\section{Introduction}
With the increase of size and functionality, networked control systems are limited by communication and energy resources \citep{zhang2001stability,hespanha2007survey,gupta2009networked,zhang2015survey}. Event-triggered communication   has received considerable attention as an effective method to mitigate the resource constraints by actively select the most valuable information according to predefined event-triggering conditions \citep{aaarzen1999simple,astrom2002comparison,heemels2012introduction}.

The scope of this work belongs to event-triggered state estimation, which aims to analyze the capacity of reconstructing the state of systems with event-triggered sensor measurements \citep{shi2016eventbook}. In event-triggered state estimation, two related agents are usually considered in networked control systems with event-triggered communication channels: one is an intelligent sensor that observes the state periodically and makes decisions as to whether to transmit the state, and the other one is the event-triggered state estimator which attempts to reconstruct the state sequentially with received information \citep{imer2005optimal,sijs2013event,shi2016eventbook}. Minimum mean square error (MMSE) estimators for Gaussian systems were designed in earlier investigations \citep{li2010event}. In \cite{wu2012event}, the measurements were sent to the observer when the system innovation was beyond a pre-set threshold, and a closed-form approximate MMSE estimator was developed. \cite{shi2014event} proposed an optimal sensor fusion approach for multiple sensors with separate event-triggering conditions under a truncated Gaussian assumption of the conditional distribution of the state. To overcome the issue caused by the Gaussian assumption of the conditional distribution, stochastic event-triggering mechanisms were designed, and corresponding event-triggered state estimators were also developed \citep{wu2014stochastic,han2015stochastic,han2017stochastic}. Later  investigations considered event-triggered communication channels with packet dropout, and the nonexistence of event-based triggers that preserve Gaussian state \citep{kung2017nonexistence} was proved, which promoted the development of event-triggered state estimation using hidden Markov models. For hidden Markov models, closed-form representations for estimators and average sensor-to-estimator communication rates were obtained using the reference measure approach \citep{shi2016event,chen2017event} for lossy communication channels. More recently, the effect of model uncertainty was considered and risk-sensitive event-triggered filtering \citep{xu2019event,huang2019robust} and smoothing \citep{cheng2021event} approaches were developed by optimizing exponential objective functions.

In networked control systems with event-triggered communication channels, sensor measurements arrive non-uniformly. To deal with this issue, several interesting investigations for the observability of linear systems with non-uniform measurements exist in the literature \citep{xu2008state,li2011state,ding2009reconstruction}.
%
%
%
For instance, a state reconstruction problem was considered for non-uniform samplings in \cite{xu2008state} for systems with binary-valued sensors. Utilizing the threshold and crossing time of the sensor, the system was proved to be observable with $n$ binary-sensor switching points for $n$th order observable continuous linear time-invariant (LTI) systems. A similar approach was adopted in \cite{li2011state}, where authors proved that if the original system is observable, the non-uniform sampled system will be observable provided that the sampling density is higher than some critical frequency, independent of the actual time sequences. In \cite{ding2009reconstruction}, the control input and the measurement output were modeled to be non-uniform through a zero-order hold from continuous-time linear systems, and the related observability issues were studied. The authors in \cite{ding2009reconstruction} pointed out that the designed non-uniform sampling pattern does not change the observability of linear systems even if the sampling interval was pathological. In \cite{ben2021time}, the observability problem for linear dynamic systems on non-uniform time domains was considered, and the duality relationship between observability and reachability was established. It is interesting to note that the observability problems considered in aforementioned investigations only considered the received measurements for state reconstruction with non-uniform sampled sensor measurements. In event-triggered state estimation, not receiving the exact sensor measurement value at a no-event instance does not equivalently mean that nothing can be inferred for the current measurement, as important information can be provided by exploiting the event-triggering conditions \citep{wu2012event,shi2014event,han2015stochastic}. This fact, however, still remain unexplored in existing studies, which motivates the investigation of this work.


In this paper, we attempt to discuss the worst-case performance of estimating the initial state of LTI systems in observability analysis for event-triggered state estimation, namely, determining a bounded region that contains the initial state, which is different from the observability aforementioned ensuring unique determination of the initial state. In order to utilize the set-valued information, a set-membership filtering approach by \cite{witsenhausen1968sets} is used to characterize a set of all possible values of the state compatible with the measurement of outputs. We consider LTI systems with bounded process disturbance and measurement noise, and the measurements of systems are sent to an estimator through a send-on-delta event-triggered communication channel. The estimator is designed using the set-membership approach \citep{durieu2001multi,yang2009set,li2020set} to obtain optimal ellipsoid approximations of system states. The main contribution of this paper is summarized as follows:

\begin{enumerate}
	\item A novel $\epsilon$-observability is defined for discrete-time LTI systems, which can evaluate how accurate we can reconstruct the initial state under the event-triggered communication strategy. The proposed $\epsilon$-observability notion focuses on the possible range determined by received measurements and information in event-triggering mechanisms at the no-event instants.
	\item A criterion is developed using a set-membership approach for the proposed $\epsilon$-observability notion, which can be used for observability analysis and observer design. A calculation approach to determinate an upper bound of parameter $\epsilon$ is provided, which can evaluate the worst-case performance of determining the initial state for LTI systems with event-triggered communication channels.
	\item An iterative set-membership observer is designed based on the proposed $\epsilon$-observability notion for LTI systems, which  consists of predictive ellipsoids inferred from the last state, ellipsoids from measurement and event-triggering conditions, and the optimal outer approximation of intersection of such two ellipsoids in the sense of trace at each step. The estimated ellipsoids are proved to be asymptotically bounded, which is verified by several numerical examples.
\end{enumerate}

The remainder of this paper is organized as follows. Section \ref{sec:prob} presents the system description, the definition of $\epsilon$-observability, and the problem formulation. A criterion to judge the $\epsilon$-observability of LTI systems and an approach to calculate an upper bound of the parameter $\epsilon$ are provided in Section \ref{sec:epobcri}. Based on the proposed $\epsilon$-observability and the developed criterion, an iterative set-membership observer is designed in Section \ref{sec:setob}. Implementation issues and numerical verification of the results are presented in Section \ref{sec:numexam}, followed by some conclusions in Section \ref{sec:conc}. 




\section{Problem Formulation}\label{sec:prob}
Conseder the following LTI system with a single output
\begin{equation}\label{eq: sys with w}
	\begin{split}
	{\bm x}_k&=A{\bm x}_{k-1}+{\bm w}_{k-1},\\
	y_k&=C{\bm x}_k+v_k,
	\end{split}
\end{equation}
where ${\bm x}_k=[x_k^1, \ldots, x_k^n]^{\rm T}\in\mathbb{R}^n$ is the state, $y_k\in\mathbb{R}$ is the measurement, and $A\in\mathbb{R}^{n\times n}, C\in\mathbb{R}^{1\times n}$ are system parameters. The process disturbance ${\bm w}_k\in\mathbb{R}^n$ and measurement noise $v_k\in\mathbb{R}$ are assumed to be two series of random variables satisfying
\begin{align}
	{\bm w}_k&\in \mathcal{W}:=\{{\bm w}:{\bm w}^{\rm T}Q^{-1}{\bm w}\leq 1\}=\mathcal{E}(0,Q),\label{eq: stQ}\\
	{ v}_k&\in \mathcal{V}:=\{{ v}:{ v}^{\rm T}R^{-1}{ v}\leq 1\}=\mathcal{E}(0,R),\label{eq: stR}
\end{align}
with $Q$ and $R$ being real symmetric positive definite matrices and $\mathcal{E}(\cdot,\cdot)$ being an ellipsoid. Note that the results obtained in this work can be extended to systems with multiple outputs, which is stated at Remark \ref{rem:extend} in Section \ref{sec:setob}. 

In this work, the measurement $y_k$ is transmitted by a deterministic event-triggered mechanism described as
\begin{equation}\label{eq: event}
	\gamma_k=\left\{\begin{array}{ll}
			1, &{\text {if}}~ (y_k-y_{\tau k})^{\rm T}\Gamma^{-1}(y_k-y_{\tau k})>1,\\
			0, &{\text {if}} ~(y_k-y_{\tau k})^{\rm T}\Gamma^{-1}(y_k-y_{\tau k})\leq1,
		\end{array}\right.
\end{equation}
where $\gamma_k=1$ denotes that the measurement $y_k$ is transmitted to the observer and $\gamma_k=0$ the otherwise, $\Gamma>0$ and $y_{\tau k}$ are known parameters. For send-on-delta mechanism \citep{miskowicz2006send,shi2014event}, $y_{\tau k}$ is the previously transmitted measurement. 

Let $\mathcal{Y}_{ok}$ denotes the set-valued information for the observer inferred based on the received measurement and event-triggering condition, which satisfies
\begin{equation}\label{eq: obinfo}
	\mathcal{Y}_{ok}:=\{y:(y-y_{\tau k})^{\rm T}M_k^{-1}(y-y_{\tau k})\leq 1\}.
\end{equation}
The matrix $M_k$ describes the information of the measurement which can be write as
\begin{equation}\label{eq: Mk}
	M_k=\left\{\begin{array}{ll}
		\Gamma, &\text{if}~ \gamma_k=0,\\
		\Gamma_e, &\text{if}~ \gamma_k=1,
	\end{array}\right.
\end{equation}
where $\Gamma_e$ is the measurement or transmission error. For example, if the measurement is transmitted by a digital wireless communication channel including an analog to digital converter (ADC) and a digital to analog converter (DAC), $\Gamma_e$ can be used to describe the quantization error of the ADC. Generally, if there is no measurement error or transmission error, $\Gamma_e$ can be set to be sufficiently small.

This work attempts to investigate the observability of system \eqref{eq: sys with w} under event-triggered mechanism \eqref{eq: event} using received information \eqref{eq: obinfo}, especially when the measurement is not transmitted. To enable our further analysis, we introduce the notion of $\epsilon$-observability as follows.
\begin{define}\label{def: epob}
	($\epsilon$-observability) Given $\epsilon>0$, the system in \eqref{eq: sys with w} is said to be $\epsilon$-observable if for any initial state ${\bm x}_0$, there exists a finite $K>0$ such that the knowledge of the measurement information $\{\mathcal{Y}_{o0},\mathcal{Y}_{o1},\ldots,\mathcal{Y}_{oK}\}$ suffices to determine a bounded ellipsoidal area which contains the initial state
	\begin{equation}\label{eq: ebobeq}
		{\bm x}_0\in\mathcal{E}(\hat{\bm x}_0,\hat{P}_0),
	\end{equation}
with ${\rm Tr}(\hat{P}_0)\leq \epsilon$. Otherwise, the system is said to be $\epsilon-$unobservable.
	\end{define}
The definition above is an extension of classical observability as in Definition 6.D2 in \cite{chen1999linear}, from uniquely determining the initial state to the determining a bounded ellipsoidal area containing the initial state. It can be observed from Definition \ref{def: epob} that $\epsilon$ is an important parameter to describe the degree of estimation accuracy of system \eqref{eq: sys with w}.

Based on the descriptions above, the objective of this work is to investigate two problems:
\begin{enumerate}
	\item How to determine the $\epsilon$-observability of LTI system \eqref{eq: sys with w}, especially the accuracy parameter $\epsilon$?
	\item How to design an iterative state estimator according to the $\epsilon$-observability of system \eqref{eq: sys with w}?
\end{enumerate}

\section{$\epsilon$-Observability Criterion}\label{sec:epobcri}
In this section, the estimation of initial state ${\bm x}_0$ utilizing the set-valued measurement information is first discussed, and then a criterion for $\epsilon$-observability of system \eqref{eq: sys with w} is developed based on the discussion. 

For convenience, several useful lemmas for ellipsoids including affine transformation, geometrical (Minkowski) sum, and intersection are stated as follows.

\begin{lem}\label{lem: affine}
	(Affine transformation) For ${\bm x}\in\mathcal{E}({\bm a},Q)$, the affine transformation $A{\bm x}+{\bm b}$ satisfy
	\begin{equation*}
		A{\bm x}+{\bm b}\in A\mathcal{E}({\bm a},Q)+b=\mathcal{E}(A{\bm a}+{\bm b}, AQA^{\rm T}).
	\end{equation*}
	\end{lem}

\begin{lem}\label{lem: sum}
	(Minkowski sum) For ${\bm x}_1\in\mathcal{E}({\bm a}_1,Q_1)$, ${\bm x}_2\in\mathcal{E}({\bm a}_2,Q_2)$, the sum ${\bm x}_1+{\bm x}_2$ satisfy
	\begin{align*}
		&{\bm x}_1+{\bm x}_2\\
		\in&\mathcal{E}({\bm a}_1,Q_1)\oplus\mathcal{E}({\bm a}_2,Q_2)\\
		\subset&\mathcal{E}({\bm a}_1+{\bm a}_2,f_+(Q_1,Q_2;p)),
	\end{align*}
where 
\begin{align*}
	f_+(Q_1,Q_2;p)&:=(1+p^{-1})Q_1+(1+p)Q_2,\\
	p&\in[\lambda_{min}^{\frac{1}{2}},\lambda_{max}^{\frac{1}{2}}],
\end{align*}
with $\lambda_{min}$ and $\lambda_{max}$ being minimal and maximal roots of the equation $\det(Q_1-\lambda Q_2)=0$, respectively.
\end{lem}

\begin{rem}\label{rem: p*}
	In Lemma \ref{lem: sum}, a set of ellipsoids containing the Minkowski sum of ellipsoids $\mathcal{E}({\bm a}_1,Q_1)$ and $\mathcal{E}({\bm a}_2,Q_2)$ are represented using parameter $p$, which can be seen as external approximations of the Minkowski sum. In this work, an optimal choice $p^*$ of parameter $p$ is utilized to minimize ${\rm Tr}(f_+(Q_1,Q_2,p))$ as
	\begin{equation}
		\begin{split}\label{eq: mintr}
		p^*&=\arg\min\limits_{p}{\rm Tr}(f_+(Q_1,Q_2,p)),\\
		{\rm s.t.}~~~p&\in\left[\lambda_{min}^{\frac{1}{2}},\lambda_{max}^{\frac{1}{2}}\right].
		\end{split}
		\end{equation}
	\end{rem}

To solve Problem \ref{eq: mintr}, let $\frac{{\rm d} {\rm Tr}(f_+(Q_1,Q_2,p))}{{\rm d} p}=0$, then we have
\begin{align*}
	\frac{{\rm d}{\rm Tr}\left((1+p^{-1})Q_1+(1+p)Q_2\right)}{{\rm d} p}&=0,\\
	{\rm d}\frac{1+p^{-1}}{{\rm d} p}{\rm Tr}(Q_1)+{\rm d}\frac{1+p}{{\rm d}p}{\rm Tr}(Q_2)&=0
\end{align*}
Thus we obtain
\begin{equation}\label{eq: p*}
		p^*=\frac{\sqrt{{\rm Tr}(Q_1)}}{\sqrt{{\rm Tr}(Q_2)}}.
\end{equation}
In this work, we denote
\begin{equation*}
	\mathcal{E}({\bm a}_1,Q_1)\oplus\mathcal{E}({\bm a}_2,Q_2)\subset\mathcal{E}({\bm a}_1+{\bm a}_2,f_+(Q_1,Q_2;p^*)),
\end{equation*}
and $f_+(Q_1,Q_2):=f_+(Q_1,Q_2;p^*)$ for simplification. The function $f_+(\cdot,\cdot)$ is used to describe the Minkowski sum of two ellipsoids, based on which the Minkowski sum of more ellipsoids can be performed successively. For example, the Minkowski sum of three ellipsoids $\mathcal{E}({\bm a}_1,Q_1), \mathcal{E}({\bm a}_2,Q_2)$, and $\mathcal{E}({\bm a}_3,Q_3)$ can be obtained as
\begin{align*}
	&\mathcal{E}({\bm a}_1,Q_1)\oplus\mathcal{E}({\bm a}_2,Q_2)\oplus\mathcal{E}({\bm a}_3,Q_3)\\
	\subset&\mathcal{E}({\bm a}_1+{\bm a}_2+{\bm a}_3,f_+(f_+(Q_1,Q_2),Q_3)).
\end{align*}
Then we define 
\begin{align*}
	f_+^{[3]}(\cdot,\cdot,\cdot)&:=f_+\left(f_+(\cdot,\cdot),\cdot\right),\\
	f_+^{[i]}(\cdots)&:=f_+\left(f_+^{[i-1]}(\cdots),\cdot\right).
\end{align*}

\begin{lem}\label{lem: intersection}
	(Intersection) For ellipsoids $\mathcal{E}({\bm a}_1,Q_1)$ and $\mathcal{E}({\bm a}_2,Q_2)$, the intersection $\mathcal{E}({\bm a}_1,Q_1) \cap\mathcal{E}({\bm a}_2,Q_2)$ satisfies
	\begin{align*}
		&\mathcal{E}({\bm a}_1,Q_1) \cap\mathcal{E}({\bm a}_2,Q_2)\\
		\subset&\mathcal{E}\left(M{\bm a}_1,MQ_1M^{\rm T}\right)\\
		&~~\oplus\mathcal{E}\left((I-M){\bm a}_2,(I-M)Q_2(I-M)^{\rm T}\right),
	\end{align*}
where $M$ is a matrix parameter with appropriate dimension.
\end{lem}

\begin{rem}
	Lemma \ref{lem: intersection} proposes an external ellipsoidal estimation of intersection $\mathcal{E}({\bm a}_1,Q_1) \cap\mathcal{E}({\bm a}_2,Q_2)$ with matrix parameter $M$. In this work, an optimal choice $M^*$ is chosen to minimize the trace of the ellipsoid as
	\begin{align}\label{eq: Mk*}
		M^*=&\arg\min\limits_{M}{\rm Tr}\left(MQ_1M^{\rm T}\right)\notag\\
		&~~+{\rm Tr}\left((I-M)Q_2(I-M)^{\rm T}\right),
		\end{align}
which can be solved according to the equation 
\begin{equation*}
	\frac{{\rm d}{\rm Tr}\left(MQ_1M^{\rm T}\right)}{{\rm d} M}+\frac{{\rm d}{\rm Tr}\left((I-M)Q_2(I-M)^{\rm T}\right)}{{\rm d} M}=0.
\end{equation*}
By direct derivation, the optimal choice $M^*$ of Problem \eqref{eq: Mk*} is
\begin{equation*}
	M^*=(Q_1+Q_2)^{-1}Q_2.
\end{equation*}
	\end{rem}

Based on the lemmas above, the $\epsilon$-observability of system \eqref{eq: sys with w} is investigated by analyzing the relationship between initial state ${\bm x}_0$ and set-valued measurement information $\{\mathcal{Y}_{o0},\mathcal{Y}_{o1},\ldots\}$ as follows.

For time instant $i>0$, the output of system \eqref{eq: sys with w} can be written as
\begin{equation}\label{eq: yi}
	y_i=CA^i{\bm x}_0+\sum\limits_{j=0}^{i-1}CA^{i-j-1}{\bm w}_j+v_i.
\end{equation}
After defining that for $i=0$, the item $\sum\limits_{j=0}^{i-1}CA^{i-j-1}{\bm w}_j$ equals $0$, equation \eqref{eq: yi} works for $i\in\left\{0,1,\ldots\right\}$. Then equivalently, we have
\begin{equation}\label{eq: CAi}
	CA^i{\bm x}_0=y_i-\sum\limits_{j=0}^{i-1}CA^{i-j-1}{\bm w}_j-v_i.
\end{equation}
Note that for $\forall {\bm w}\in\mathcal{W}$ and $\forall v\in\mathcal{V}$, the corresponding negative elements satisfy $-{\bm w}\in\mathcal{W}$ and $-v\in\mathcal{V}$, respectively. Thus according to the definition of Minkowski sum stated in Lemma \ref{lem: sum}, equation \eqref{eq: CAi} leads to 
\begin{align}
	&CA^i{\bm x}_0\notag\\
	\in&\mathcal{Y}_{oi}\bigoplus\limits_{j=0}^{i-1}CA^{i-j-1}\mathcal{W}\oplus\mathcal{V}\notag\\
	=&\mathcal{E}\left(y_{\tau i},f_+^{[i+1]}(M_i,CA^{i-1}Q(A^{i-1})^{\rm T}C^{\rm T},\ldots,CQC^{\rm T},R)\right)\notag\\
	=&\!\!:\mathcal{E}(y_{\tau i},W_{0,i}).\label{eq: defW}
\end{align}

By defining the Minkowski sum of several ellipsoids in equation \eqref{eq: defW}, the corresponding inequalities can be written as
\begin{align}\label{eq: ineqW}
	(CA^{i}{\bm x}_0-y_{\tau i})^{\rm T}W_{0,i}^{-1}(CA^{i}{\bm x}_0-y_{\tau i})\leq 1.
\end{align}

In order to combine several inequalities together, we introduce $a=[a_0,\ldots, a_{n-1}]$ satisfying
\begin{equation}\label{eq: defa}
	\sum\limits_{i=0}^{n-1} a_i=1, a_i>0, i\in\{0,1,\ldots, n-1\}.
\end{equation}
Then by multiplying $a_i$, inequality \eqref{eq: ineqW} turns to
\begin{equation}\label{eq: ineqaW}
(CA^{i}{\bm x}_0-y_{\tau i})^{\rm T}a_i W_{0,i}^{-1}(CA^{i}{\bm x}_0-y_{\tau i})\leq a_i.
\end{equation} 

For $i\in\{0,\ldots,n-1\}$, after defining
\begin{align}
	O&:=\left[\begin{array}{c}
		C\\
		CA\\
		\vdots\\
		CA^{n-1}
	\end{array}\right], 
~~~Y_{\tau 0}:=\left[\begin{array}{c}
	y_{\tau 0}\\
	y_{\tau 1}\\
	\vdots\\
	y_{\tau n-1}
\end{array}\right],\label{eq: O}\\
Q_0(a)&:=\left[\begin{array}{cccc}
	a_0W_{0,0}^{-1} & 0& \cdots&0\\
	0& a_1W_{0,1}^{-1}&\cdots&0\\
	\vdots&\vdots&\ddots&\vdots\\
	0&0&\cdots&a_{n-1}W_{0,n-1}^{-1}
\end{array}\right],\notag
\end{align}
with $Q_0(a)$ being a diagonal matrix, sum of inequalities as form \eqref{eq: ineqaW} can be written as
\begin{align}
	\sum\limits_{i=0}^{n-1}(CA^{i}{\bm x}_0-y_{\tau i})^{\rm T}a_i W_{0,i}^{-1}(CA^{i}{\bm x}_0-y_{\tau i})&\leq	\sum\limits_{i=0}^{n-1} a_i\notag\\
	(O {\bm x}_0-Y_{\tau 0})^{\rm T}Q_0(a)(O {\bm x}_0-Y_{\tau 0})&\leq 1, \label{eq: sumineq} 
	\end{align}
and the matrix $O$ is called observability matrix.

Based on the inequalities aforementioned, the $\epsilon$-observability criterion of system \eqref{eq: sys with w} is stated as Theorem \ref{thm: epob}.

 \begin{thm}\label{thm: epob}
    An LTI system in \eqref{eq: sys with w} is $\epsilon$-observable if matrix $O$ defined in equation \eqref{eq: O} has full column rank and $K=n-1$. The parameter $\epsilon$ satisfies
    \begin{equation*}
    	\epsilon=\max\limits_{\gamma_i} {\rm Tr}\left(O^{-1}Q_0^{-1}(a)(O^{\rm T})^{-1}\right), i\in\{0,\ldots,n-1\}.
    \end{equation*}
 	\end{thm}
\begin{pf}
	If the matrix $O\in\mathbb{R}^{n\times n}$ is column full rank, the inequality \eqref{eq: sumineq} can be rewritten as
	\begin{align*}
		\left[O({\bm x}_0-O^{-1}Y_{\tau 0})\right]^{\rm T}Q_0(a)\left[O({\bm x}_0-O^{-1}Y_{\tau 0})\right]&\leq 1\\
		({\bm x}_0-O^{-1}Y_{\tau 0})^{\rm T}\left[O^{\rm T}Q_0(a)O\right]({\bm x}_0-O^{-1}Y_{\tau 0})&\leq 1.
	\end{align*}
Thus we obtain an ellipsoidal estimation of the initial state ${\bm x}_0$ as
\begin{equation}\label{eq: x0est}
	{\bm x}_0\in\mathcal{E}(O^{-1}Y_{\tau 0},O^{-1}Q_0^{-1}(a)(O^{\rm T})^{-1}).
\end{equation}
Let $\hat{\bm x}_0=O^{-1}Y_{\tau 0}$, and $\hat{P}_0=O^{-1}Q_0^{-1}(a)(O^{\rm T})^{-1}$, then equation \eqref{eq: ebobeq} is satisfied. Observing the definition of $W_{0,i}$ in equation \eqref{eq: defW}, it can be find that $W_{0,i}$ is defined based on the Minkowski sum of several ellipsoids, where $M_i$ is the only uncertain matrix. The definition of $M_i$ in equation \eqref{eq: Mk} shows that $M_i$ equals either $\Gamma$ or $\Gamma_e$, which is determined by $\gamma_i$. Thus for a fixed parameter $a$, there are $2^n$ possible values for the matrix $Q_0(a)\in\mathbb{R}^{n\times n}$. After defining $\epsilon$ as
    \begin{equation*}
	\epsilon:=\max\limits_{\gamma_i} {\rm Tr}\left(O^{-1}Q_0^{-1}(a)(O^{\rm T})^{-1}\right), i\in\{0,\ldots,n-1\},
\end{equation*}
we have ${\rm Tr}(\hat{P}_0)\leq \epsilon$.

According to the Definition \ref{def: epob}, the $\epsilon$-observability of system \eqref{eq: sys with w} is proved.\hfill\QEDopen
	\end{pf}

\begin{rem}
	Theorem \ref{thm: epob} proposes a criterion of the $\epsilon$-observability for system in \eqref{eq: sys with w}. The parameter $\epsilon$ is an indicator to evaluate the uncertainty on estimating the initial state ${\bm x}_0$ inferred based on the received measurement and the knowledge of the event-triggered communication channel. Note that the parameter $\epsilon$ is determined by the characteristic of the system $A, C, Q, R$, the parameter of the communication channel $\Gamma, \Gamma_e$, and the designed parameter $a$. The parameter $\epsilon$ does not concern the specific communication condition ${\rm \gamma}$, which reflects the worst case of determining the region of the initial state.
	\end{rem}

\section{Event-triggered Set-membership Observer}\label{sec:setob}
This section aims to design an estimator for system \eqref{eq: sys with w}, which could calculate an external ellipsoidal approximation of state ${\bm x}_k$ iteratively. This section consists of four parts: analysis of predictive possible region of ${\bm x}_k$ based on the estimated region of ${\bm x}_{k-1}$ (prior set), analysis of the possible region of ${\bm x}_k$ inferred from set-valued measurement information (measurement information set), combination of prior set and measurement information set (iterative state estimator), and the analysis of the convergence of the proposed method.

\subsection{Prior Set}\label{subsec: prio}
Assume that at time instant $k$, we have obtained an ellipsoidal approximation for ${\bm x}_{k-1}$ as
\begin{equation}
	{\bm x}_{k-1}\in\hat{X}_{k-1}:=\mathcal{E}(\hat{\bm x}_{k-1},\hat{P}_{k-1}),
\end{equation}  
where $\hat{X}_{k-1}$ is an estimated ellipsoid with parameters $\hat{\bm x}_{k-1}$ and $\hat{P}_{k-1}$. Then according to Lemma \ref{lem: affine}, we have
\begin{equation}
	A{\bm x}_{k-1}\in\mathcal{E}(A\hat{\bm x}_{k-1},A\hat{P}_{k-1}A^{\rm T}).
\end{equation}
Considering the disturbance ${\bm w}_{k-1}\in\mathcal{W}$, the prior estimation of ${\bm x}_k$ can be written as
\begin{align}
	{\bm x}_{k}&=A{\bm x}_{k-1}+{\bm w}_{k-1}\notag\\
	&\in\mathcal{E}(A\hat{\bm x}_{k-1},A\hat{P}_{k-1}A^{\rm T})\oplus\mathcal{W}\notag\\
	&\subset\mathcal{E}(A{\bm x}_{k-1},f_+(A\hat{P}_{k-1}A^{\rm T},Q)).
\end{align}
After defining
\begin{align}
	\check{P}_k
	:=&f_+(A\hat{P}_{k-1}A^{\rm T},Q)\notag\\
	=&\left(1+\sqrt{\frac{{\rm Tr} (Q)}{{\rm Tr}(A\hat{P}_{k-1}A^{\rm T})}}\right)A\hat{P}_{k-1}A^{\rm T}\notag\\
	&~~+\left(1+\sqrt{\frac{{\rm Tr}(A\hat{P}_{k-1}A^{\rm T})}{{\rm Tr} (Q)}}\right)Q,\label{eq: pcheck}
\end{align}
we obtain a predictive prior set estimation $\check{X}_k$ for state ${\bm x}_k$ as
\begin{equation}\label{eq: xkqian}
	{\bm x}_{k}\in \check{X}_k:=\mathcal{E}(A\hat{\bm x}_{k-1},\check{P}_{k}).
\end{equation}

\subsection{Measurement information Set}\label{subsec: meas}
The analysis of the measurement information set is similar to the analysis of $\epsilon$-observability in equations \eqref{eq: yi}-\eqref{eq: x0est}, and thus is stated in brief as follows.

Consider $n$ measurements from time instant $k$, i.e., $y_{k+i}, i\in\{0,1,\ldots, n-1\}$. The measurements satisfy
\begin{equation}
	y_{k+i}=CA^i{\bm x}_k+\sum\limits_{j=0}^{i-1}CA^{i-j-1}{\bm w}_{k+j}+v_{k+i}.
\end{equation}
Then similar to equation \eqref{eq: defW}, we have
\begin{align}
	&CA^{i}{\bm x}_k\notag\\
	=&y_{k+i}-\sum\limits_{j=0}^{i-1}CA^{i-j-1}{\bm w}_{k+j}-v_{k+i}\notag\\
	\in&\mathcal{Y}_{o(k+i)}\bigoplus\limits_{j=0}^{i-1}CA^{i-j-1}\mathcal{W}\oplus\mathcal{V}\notag\\
	\subset&\mathcal{E}(y_{\tau(k+i)},W_{k,i}),\label{eq: defWki}
\end{align}
where 
\begin{align*}
&	W_{k,i}\\:=&f_+^{[i+1]}(M_{k+i},CA^{i-1}Q(A^{i-1})^{\rm T}C^{\rm T},\ldots,CQC^{\rm T},R),
\end{align*}
and $M_{k+i}$ can be determined according to equation \eqref{eq: Mk}. The ellipsoid in \eqref{eq: defWki} can be rewritten as
\begin{equation*}
	(CA^i{\bm x}_k-y_{\tau (k+i)})^{\rm T}W_{k,i}^{-1}(CA^i{\bm x}_k-y_{\tau (k+i)})\leq 1.
\end{equation*}
Then we have
\begin{equation}
	(O{\bm x}_k-Y_{\tau k})^{\rm T}Q_k(a)(O{\bm x}_k-Y_{\tau k})\leq1,
\end{equation}
where $a$, $O$ are defined in equations \eqref{eq: defa} and \eqref{eq: O} respectively, and $Y_{\tau k}, Q_k(a)$ are defined as follows:
\begin{align}
	Y_{\tau k}&:=\left[\begin{array}{c}
		y_{\tau k}\\
		y_{\tau k+1}\\
		\vdots\\
		y_{\tau k+n-1}
	\end{array}\right],\label{eq: Ytk}\\
	Q_k(a)&:=\left[\begin{array}{cccc}
		a_0W_{k,0}^{-1} & 0& \cdots&0\\
		0& a_1W_{k,1}^{-1}&\cdots&0\\
		\vdots&\vdots&\ddots&\vdots\\
		0&0&\cdots&a_{n-1}W_{k,n-1}^{-1}
	\end{array}\right].\label{eq: Qka}
	\end{align}
For invertible observability matrix $O$, we have
\begin{equation*}
	({\bm x}_k-O^{-1}Y_{\tau k})^{\rm T}O^{\rm T}Q_k(a)O({\bm x}_k-O^{-1}Y_{\tau k})\leq 1.
\end{equation*}
Then a set-valued estimation $\bar{X}_{k}$ for ${\bm x}_k$ can be obtained as
\begin{equation}\label{eq: xkhou}
	{\bm x}_k\in\bar{X}_k:=\mathcal{E}(O^{-1}Y_{\tau k},\bar{P}_k),
\end{equation}
where
\begin{equation}\label{eq: OQK}
	\bar{P}_k=O^{-1}(Q_k(a))^{-1}(O^{\rm T})^{-1}.
\end{equation}

\begin{rem}
	The matrix $\bar{P}_k$ can be calculated utilizing two matrices, $O$ and $Q_k(a)$, where $O$ is the observability matrix as a constant for system \eqref{eq: sys with w}. According to equation \eqref{eq: defWki}, $W_{k,i}$ can be calculated by matrices $C, A, Q, R$ and $M_{k+i}$. Since there are two possible values for $M_{k+i}, i\in\{0,1,\ldots,n-1\}$, namely, $\Gamma$ and $\Gamma_e$, $Q_k(a)$ has $2^n$ possible values, which involves to $W_{k,0}, W_{k,1}, \ldots, W_{k,n-1}$. The specific value of $Q_k(a)$ is determined by ${\bm \gamma}=[\gamma_k,\ldots,\gamma_{k+n-1}]$. Compare the trace of all possible values of matrix $\bar{P}_k$ and write the largest one as 
	\begin{equation*}
		\max\limits_{{\bm \gamma}}{\rm Tr}(\bar{P}_k),
	\end{equation*}
	which is useful at the analysis of the asymptotic property of the matrix ${\rm Tr}(\hat{P}_k)$.
	\end{rem}

\subsection{Iterative State Estimation}\label{subsec: state est}
It can be observed from the results of the subsections above that two ellipsoids containing the state ${\bm x}_k$ are determined based on the estimation at last instant $\hat{X}_{k-1}$ and set-valued measurement information $\mathcal{Y}_{o(k+i)},i\in\{0,1,\ldots,n-1\}$, respectively. Obviously, the state ${\bm x}_k$ belongs to the intersection of sets $\bar{X}_k$ and $\check{X}_k$ as
\begin{equation*}
	\left.\begin{array}{l}
			{\bm x}_k\in\bar{X}_k\\
			{\bm x}_k\in\check{X}_k
		\end{array}\right\}
	\Rightarrow{\bm x}_k\in\bar{X}_k\cap\check{X}_k.
\end{equation*}
To design an iterative state estimator, the external ellipsoidal estimation of the intersection $\bar{X}_k\cap\check{X}_k$ can be obtained as
\begin{align}
	{\bm x}_k\in&\bar{X}_k\cap\check{X}_k\notag\\
	=&\mathcal{E}(O^{-1}Y_{\tau k},\bar{P}_k)\cap\mathcal{E}(A\hat{\bm x}_{k-1},\check{P}_k)\notag\\
	\subset&\mathcal{E}(M_kO^{-1}Y_{\tau k},M_k\bar{P}_kM_k^{\rm T})\label{eq: interbarc}\\
	&\;\;\oplus \mathcal{E}\left((I-M_k)A\hat{\bm x}_{k-1},(I-M_k)\check{P}_k(I-M_k)^{\rm T}\right)\notag,
\end{align}
where equation \eqref{eq: interbarc} is obtained according to Lemma \ref{lem: intersection}. By minimizing the trace of the external ellipsoidal estimation as
\begin{equation*}
	M^*_k=\arg\min\limits_{M_k} {\rm Tr}(M_k\bar{P}_kM_k^{\rm T})+{\rm Tr}((I-M_k)\check{P}_k(I-M_k)^{\rm T}),
\end{equation*}
the optimal parameter $M^*_k$ can be obtained according to equation \eqref{eq: Mk*} as
\begin{equation}\label{eq: Mk*use}
	M_k^*=(\bar{P}_k+\check{P}_k)^{-1}\check{P}_k.
\end{equation}
Then equation \eqref{eq: interbarc} turns to 
\begin{align}
	{\bm x}_k\in&\mathcal{E}(M_k^*O^{-1}Y_{\tau k},M_k^*\bar{P}_k(M_k^*)^{\rm T})\notag\\
	&~~~~\oplus \mathcal{E}\left((I-M_k^*)A\hat{\bm x}_{k-1},(I-M_k^*)\check{P}_k(I-M_k^*)^{\rm T}\right)\notag\\
	\subset&\mathcal{E}\Big(M_k^*O^{-1}Y_{\tau k}+(I-M_k^*)A\hat{\bm x}_{k-1},\label{eq: summin}\\
	&~~~~f_+(M_k^*\bar{P}_k(M_k^*)^{\rm T},(I-M_k^*)\check{P}_k(I-M_k^*)^{\rm T})\Big).\notag
\end{align}
Note that in equation \eqref{eq: summin}, the Minkowski sum of two ellipsoids is denoted by function $f_+(\cdot,\cdot)$, which minimizes the trace of the sum according to Remark \ref{rem: p*}. In fact, the meaning expressed by the equation \eqref{eq: summin} is
\begin{align*}
	{\bm x}_k\in&\mathcal{E}\Big(M_k^*O^{-1}Y_{\tau k}+(I-M_k^*)A\hat{\bm x}_{k-1},\\
	&~~~~f_+(M_k^*\bar{P}_k(M_k^*)^{\rm T},(I-M_k^*)\check{P}_k(I-M_k^*)^{\rm T};p^*)\Big),
\end{align*}
where $p^*$ is obtained according to Remark \ref{rem: p*} and equation \eqref{eq: p*} as
\begin{align}
	p^*\!\!&=\!\!\arg\min\limits_p{\rm Tr}(f_+(M_k^*\bar{P}_k(M_k^*)^{\rm T}\!,\!(I\!\!-\!\!M_k^*)\check{P}_k(I\!\!-\!\!M_k^*)^{\rm T}\!;\!p)\!)\notag\\
	&=\frac{\sqrt{{\rm Tr}(M_k^*\bar{P}_k(M_k^*)^{\rm T})}}{\sqrt{{\rm Tr}((I-M_k^*)\check{P}_k(I-M_k^*)^{\rm T})}}.\label{eq: p*tr}
	\end{align}
Write
\begin{align}
	\hat{\bm x}_k&=M_k^*O^{-1}Y_{\tau k}+(I-M_k^*)A\hat{\bm x}_{k-1},\\
	\hat{P}_k&:=f_+(M_k^*\bar{P}_k(M_k^*)^{\rm T},(I-M_k^*)\check{P}_k(I-M_k^*)^{\rm T}).\label{eq: hatP}
\end{align}
Then an alternative state estimation approach is proposed in Algorithm \ref{alg: filter}.

\begin{algorithm}[htp]
	\caption{Alternative set-membership state estimation.}\label{alg: filter}
	\begin{algorithmic}[1]
		\STATE Input measurement information $\mathcal{Y}_{ok}, k\in\{0,1,\ldots,N\}$ and parameters $A, C, Q, R$;\\
		\STATE Calculate observability matrix $O$ according to equation \eqref{eq: O};\\
		\FOR {$k\in\{0,1,\ldots,N-(n-1)\}$}
		\STATE Calculate $Y_{\tau k}$ and $Q_k(a)$ according to equations \eqref{eq: Ytk} and \eqref{eq: Qka}, respectively;\\
		\STATE $\bar{P}_k=(O^{\rm T}Q_k(a)O)^{-1}$;\\
		\STATE $\bar{X}_k=\mathcal{E}(O^{-1}Y_{\tau k},\bar{P}_k)$;
		\IF {$k=0$}
		\STATE $\hat{X}_k=\bar{X}_k$;\\
		\ELSE 
		\STATE $\check{P}_k=f_+(A\hat{P}_{k-1}A^{\rm T},Q)$;\\
		\STATE $\check{X}_k=\mathcal{E}(A{\bm x}_{k-1},\check{P}_k)$;\\
		\STATE $M_k^*=(\bar{P}_k+\check{P}_k)^{-1}\check{P}_k$;\\
		\STATE $p^*=\frac{\sqrt{{\rm Tr}(M_k^*\bar{P}_k(M_k^*)^{\rm T})}}{\sqrt{{\rm Tr}((I-M_k^*)\check{P}_k(I-M_k^*)^{\rm T})}}$;\\
		\STATE $\hat{P}_k=f_+(M_k^*\bar{P}_k(M_k^*)^{\rm T},(I-M_k^*)\check{P}_k(I-M_k^*)^{\rm T})$;\\
		\STATE $\hat{\bm x}_k=M_k^*O^{-1}Y_{\tau k}+(I-M_k^*)A\hat{\bm x}_{k-1}$;\\
		\STATE $\hat{X}_k=\mathcal{E}(\hat{\bm x}_k,\hat{P}_k)$;\\
		\ENDIF 
		\ENDFOR 
		\STATE Output set-membership estimations $\hat{X}_k, k\in\{0,1,\ldots,N-n+1\}$.
	\end{algorithmic}
\end{algorithm}

In Algorithm \ref{alg: filter}, the observability matrix $O$ is first calculated (Line 2), based on which the $\epsilon$-observability of the system can be determined according to Theorem \ref{thm: epob}. Then for every time instant $k$, the matrices $Y_{\tau k}$ and $Q_k(a)$ are calculated based on the set-valued measurement information received by the estimator (Line 4). After calculating the measurement information set $\bar{X}_k$ (Lines 5-6) and prior set $\check{X}_k$ (Lines 10-11), the optimal external ellipsoidal estimation of the intersection $\hat{X}_k\supset\bar{X}_k\cap\check{X}_k$ is calculated (Lines 12-16).

\begin{rem}
	In Algorithm \ref{alg: filter}, a set-membership state estimator is proposed with some time delay. By `time delay' we mean that, the measurement information $\{k,k+1,\ldots,k+n-1\}$ is required to obtain the set-membership estimation for state ${\bm x}_k$. At time instant $k+n-1$, utilizing the affine transformation and external estimation of the Minkowski sum stated in Lemmas \ref{lem: affine} and \ref{lem: sum}, a predictive region can be calculated for ${\bm x}_{k+i}, i\in\{1,\ldots\,n-1\}$ as
	\begin{equation}
		{\bm x}_{k+i}\in\mathcal{E}(A^i\hat{\bm x}_k,\tilde{P}_{k+i}), i\in\{1,\ldots,n+1\},
		\end{equation}
	where $\tilde{P}_{k+i}$ satisfies
	\begin{align}
		\tilde{P}_{k+1}&=f_+(A\hat{P}_kA^{\rm T},Q),\label{eq: pt1}\\
		\tilde{P}_{k+i}&=f_+(A\tilde{P}_{k+i-1}A^{\rm T},Q),~i\in\{2,\ldots,n-1\}\}.\label{eq: ptk}
	\end{align}
Combining Algorithm \ref{alg: filter} and equations \eqref{eq: pt1}-\eqref{eq: ptk}, we have developed a set-membership state estimator without time delay.
\end{rem}

\subsection{Convergence Analysis}
In the SubSection \ref{subsec: state est}, an ellipsoidal set-membership state estimator is developed in Algorithm \ref{alg: filter} by combining the measurement information set $\bar{X}_k$ proposed in Section \ref{subsec: meas} and the prior set $\check{X}_{k}$ proposed in Section \ref{subsec: prio}. This subsection focuses on the convergence of the proposed algorithm, which can be analyzed by investigating the asymptotic property of the posterior set $\lim\limits_{k\rightarrow+\infty}{\rm Tr}(\hat{P}_k)$.

In this subsection, the trace variation of the matrix $\hat{P}_k$ is firstly derived. Then two inequalities are proved, which is useful for analyzing the asymptotic property of the posterior set. At the end of this subsection, $\lim\limits_{k\rightarrow+\infty}{\rm Tr}(\hat{P}_k)$ is proved to be bounded and an upper bound is given.

From equation \eqref{eq: hatP}, the trace of $\hat{P}_k$ can be written as
\begin{align}
	&{\rm Tr}(\hat{P}_k)\notag\\
	=&{\rm Tr}\left(f_+(M_k^*\bar{P}_k(M_k^*)^{\rm T},(I-M_k^*)\check{P}_k(I-M_k^*)^{\rm T})\right)\notag\\
	=&{\rm Tr}((1+\frac{1}{p^*})M^*_k\bar{P}_k(M^*_k)^{\rm T})\notag\\
	&~~+{\rm Tr}((1+p^*)(I-M^*_k)\check{P}_k(I-M_k^*)^{\rm T}).\label{eq: Tr1}
\end{align}
\begin{figure*}[ht] 
	\begin{align}
		&{\rm Tr}((1+\frac{1}{p^*})M^*_k\bar{P}_k(M^*_k)^{\rm T})+{\rm Tr}((1+p^*)(I-M^*_k)\check{P}_k(I-M_k^*)^{\rm T})\notag\\
		=&\frac{\sqrt{{\rm Tr}(M_k^*\bar{P}_k(M_k^*)^{\rm T})}+\sqrt{{\rm Tr}((I-M_k^*)\check{P}_k(I-M_k^*)^{\rm T})}}{\sqrt{{\rm Tr}(M_k^*\bar{P}_k(M_k^*)^{\rm T})}}{\rm Tr}(M_k^*\bar{P}_k(M_k^*)^{\rm T})\notag\\
		&~~+\frac{\sqrt{{\rm Tr}(M_k^*\bar{P}_k(M_k^*)^{\rm T})}+\sqrt{{\rm Tr}((I-M_k^*)\check{P}_k(I-M_k^*)^{\rm T})}}{\sqrt{{\rm Tr}((I-M_k^*)\check{P}_k(I-M_k^*)^{\rm T})}}{\rm Tr}((I-M_k^*)\check{P}_k(I-M_k^*)^{\rm T})\notag\\
		=&\left(\sqrt{{\rm Tr}(M_k^*\bar{P}_k(M_k^*)^{\rm T})}+\sqrt{{\rm Tr}((I-M_k^*)\check{P}_k(I-M_k^*)^{\rm T})}\right)^2\label{eq: sqsq2}
	\end{align}
	\hrulefill
\end{figure*}
Substitute $p^*$ in equation \eqref{eq: p*tr} into equation \eqref{eq: Tr1}, then the relationship between $\hat{P}_k$, $\bar{P}_k$, and $\check{P}_k$ is stated in equation \eqref{eq: sqsq2}, which can be rewritten as
\begin{align}
	\sqrt{{\rm Tr}(\hat{P}_k)}=&\sqrt{{\rm Tr}(M_k^*\bar{P}_k(M_k^*)^{\rm T})}\notag\\
	&~~+\sqrt{{\rm Tr}((I-M_k^*)\check{P}_k(I-M_k^*)^{\rm T})}.\label{eq: sqsqsq}
\end{align}
Noting that $\check{P}_k=f_+(A\hat{P}_{k-1}A^{\rm T},Q)$, the relationship between $\hat{P}_k$ and $\hat{P}_{k-1}$ is implied in equation \eqref{eq: sqsqsq}, which is difficult to analyze. Thus two inequalities are proposed to scale equation \eqref{eq: sqsqsq}, which are proved by a constructive method as follows.

\begin{lem}\label{lem: ineq}
	Terms in equation \eqref{eq: sqsqsq} satisfy
	\begin{align}
		\sqrt{{\rm Tr}(M_k^*\bar{P}_k(M_k^*)^{\rm T})}&\leq\sqrt{{\rm Tr}(\bar{P}_k)},\label{eq: scal1}\\
		\sqrt{{\rm Tr}((I-M_k^*)\check{P}_k(I-M_k^*)^{\rm T})}&\leq\sqrt{{\rm Tr}(\check{P}_k)},\label{eq: scal2}
	\end{align}
	where $M_k^*$ is given by equation \eqref{eq: Mk*use}.
	\end{lem}
\begin{pf}
	The proof of inequality \eqref{eq: scal1} is firstly stated as follows.
	Since $\bar{P}_k$ and $\check{P}_k$ are symmetric positive semi-definite, the products of them $\bar{P}_k\bar{P}_k\bar{P}_k^{\rm T}$, $\check{P}_k\bar{P}_k\bar{P}_k^{\rm T}$ and $\bar{P}_k\bar{P}_k\check{P}_k^{\rm T}$ are positive semi-definite, i.e.,
	\begin{equation*}
		\bar{P}_k\bar{P}_k\bar{P}_k^{\rm T}+\bar{P}_k\bar{P}_k\check{P}_k^{\rm T}+\check{P}_k\bar{P}_k\bar{P}_k^{\rm T}\geq0,
	\end{equation*}
and thus 
	\begin{align*}
		\bar{P}_k\bar{P}_k\bar{P}_k^{\rm T}+\bar{P}_k\bar{P}_k\check{P}_k^{\rm T}+\check{P}_k\bar{P}_k\bar{P}_k^{\rm T}+\check{P}_k\bar{P}_k\check{P}_k^{\rm T}~~~~~~~\\-\check{P}_k\bar{P}_k\check{P}_k^{\rm T}\geq0\\
		(\bar{P}_k+\check{P}_k)\bar{P}_k(\bar{P}_k+\check{P}_k)^{\rm T}-\check{P}_k\bar{P}_k\check{P}_k^{\rm T}\geq 0.
	\end{align*}
	Since the congruent transformation does not change the positive exponential inertial and the negative exponential inertial of a matrix, we obtain
	\begin{align*}
		&(\bar{P}_k+\check{P}_k)^{-1}(\bar{P}_k+\check{P}_k)\bar{P}_k(\bar{P}_k+\check{P}_k)^{\rm T}((\bar{P}_k+\check{P}_k)^{\rm T})^{-1}\\
		&~~-(\bar{P}_k+\check{P}_k)^{-1}\check{P}_k\bar{P}_k\check{P}_k^{\rm T}((\bar{P}_k+\check{P}_k)^{\rm T})^{-1}\geq 0,
	\end{align*}
and
\begin{align}
	(\bar{P}_k+\check{P}_k)^{-1}\check{P}_k\bar{P}_k\check{P}_k^{\rm T}((\bar{P}_k+\check{P}_k)^{\rm T})^{-1}-\bar{P}_k&\leq 0\notag\\
	M^*_k\bar{P}_k(M^*_k)^{\rm T}-\bar{P}_k&\leq 0.\label{eq: mid1}
\end{align}
According to trace property of matrices ${\rm Tr}(\cdot)$, the inequality in \eqref{eq: mid1} leads to 
\begin{align*}
	{\rm Tr}(	M^*_k\bar{P}_k(M^*_k)^{\rm T})-{\rm Tr}(\bar{P}_k)\leq 0,\\
	\sqrt{{\rm Tr}(	M^*_k\bar{P}_k(M^*_k)^{\rm T})}\leq\sqrt{{\rm Tr}(\bar{P}_k)},
\end{align*}
which completes the proof of inequality \eqref{eq: scal1}.

Then the proof of the inequality \eqref{eq: scal2} can be performed similar to the proof of the inequality \eqref{eq: scal1}, which is briefly stated as follows.
According to equation \eqref{eq: Mk*use}, we have
\begin{align}
	&(I-M_k^*)\notag\\
	=&(\bar{P}_k+\check{P}_k)^{-1}(\bar{P}_k+\check{P}_k)-(\bar{P}_k+\check{P}_k)^{-1}\check{P}_k\notag\\
	=&(\bar{P}_k+\check{P}_k)^{-1}\bar{P}_k.\label{eq: mid3}
\end{align}
	 The symmetric positive semi-definiteness of matrices $\bar{P}_k$ and $\check{P}_k$ leads to
	 \begin{equation}\label{eq: mid2}
	 	\bar{P}_k\check{P}_k\check{P}_k^{\rm T}+\check{P}_k\check{P}_k\bar{P}_k^{\rm T}+\check{P}_k\check{P}_k\check{P}_k^{\rm T}\geq0.
	 \end{equation}
By adding and subtracting $\bar{P}_k\check{P}_k\bar{P}_k^{\rm T}$ to the left side of inequality in \eqref{eq: mid2}, we have
\begin{align*}
	(\bar{P}_k+\check{P}_k)\check{P}_k(\bar{P}_k+\check{P}_k)^{\rm T}-\bar{P}_k\check{P}_k\bar{P}_k^{\rm T}&\geq 0\\
	\check{P}_k-(\bar{P}_k+\check{P}_k)^{-1}\bar{P}_k\check{P}_k\bar{P}_k^{\rm T}((\bar{P}_k+\check{P}_k)^{-1})^{\rm T}&\geq 0\\
	\sqrt{{\rm Tr}((\bar{P}_k+\check{P}_k)^{-1}\bar{P}_k\check{P}_k\bar{P}_k^{\rm T}((\bar{P}_k+\check{P}_k)^{-1})^{\rm T})}&\leq\sqrt{{\rm Tr}(\check{P}_k)}\\
	\sqrt{{\rm Tr}((I-M_k^*)\check{P}_k(I-M_k^*)^{\rm T})}&\leq\sqrt{{\rm Tr}(\check{P}_k)},
\end{align*}
where the last inequality can be obtained according to equation \eqref{eq: mid3}, and the proof of inequality \eqref{eq: scal2} is completed.\hfill\QEDopen
\end{pf}

Utilizing the scaling inequalities in Lemma \ref{lem: ineq}, the asymptotic property of the posterior set $\lim\limits_{k\rightarrow+\infty}{\rm Tr}(\hat{P}_k)$ can be analyzed based on equation \eqref{eq: sqsqsq} as the following theorem.

\begin{thm}\label{thm: trP} 
    For the system in \eqref{eq: sys with w}, if the spectral norm $\|A\|< 1$, the trace of the posterior ellipsoidal estimation set ${\rm Tr}(\hat{P}_k)$ is asymptotically bounded as
    \begin{equation*}
    	\lim\limits_{k\rightarrow+\infty}\sqrt{{\rm Tr}(\hat{P}_k)}\leq\max\limits_{\bm \gamma}\frac{\sqrt{{\rm Tr}(\bar{P})}+\sqrt{{\rm Tr}(Q)}}{1-\|A\|}.
    \end{equation*}
	\end{thm}
\begin{pf}
By substituting inequalities \eqref{eq: scal1} and \eqref{eq: scal2} into equation \eqref{eq: sqsqsq}, we have
\begin{align}
	\sqrt{{\rm Tr}(\hat{P}_k)}=&\sqrt{{\rm Tr}(M_k^*\bar{P}_k(M_k^*)^{\rm T})}\notag\\
	&~~+\sqrt{{\rm Tr}((I-M_k^*)\check{P}_k(I-M_k^*)^{\rm T})}\notag\\
	\leq&\sqrt{{\rm Tr}(\bar{P}_k)}+\sqrt{{\rm Tr}(\check{P}_k)}.\label{eq: sqpsq}
\end{align}

According to the definition of matrix $\check{P}_k$ in equation \eqref{eq: pcheck}, the trace of $\check{P}_k$ satisfies
\begin{align}
	&{\rm Tr}(\check{P}_k)\notag\\
	=&{\rm Tr}(f_+(A\hat{P}_{k-1}A^{\rm T},Q))\notag\\
	=&\frac{\sqrt{{\rm Tr}(A\hat{P}_{k-1}A^{\rm T})}+\sqrt{{\rm Tr}(Q)}}{\sqrt{{\rm Tr}(A\hat{P}_{k-1}A^{\rm T})}}{\rm Tr}(A\hat{P}_{k-1}A^{\rm T})\notag\\
	&~~+\frac{\sqrt{{\rm Tr}(A\hat{P}_{k-1}A^{\rm T})}+\sqrt{{\rm Tr}(Q)}}{\sqrt{{\rm Tr}(Q)}}{{\rm Tr}(Q)}\notag\\
	=&{\rm Tr}(A\hat{P}_{k-1}A^{\rm T})+{\rm Tr}(Q)+2\sqrt{{\rm Tr}(A\hat{P}_{k-1}A^{\rm T})}\sqrt{{\rm Tr}(Q)}\notag\\
	=&\left(\sqrt{{\rm Tr}(A\hat{P}_{k-1}A^{\rm T})}+\sqrt{{\rm Tr}(Q)}\right)^2.\label{eq: sqsq22}
\end{align}
After combining inequality \eqref{eq: sqpsq} and equation \eqref{eq: sqsq22}, we obtain
\begin{align*}
	\sqrt{{\rm Tr}(\hat{P}_k)}&=\sqrt{{\rm Tr}(A\hat{P}_{k-1}A^{\rm T})}+\sqrt{{\rm Tr}(\bar{P}_k)}+\sqrt{{\rm Tr}(Q)}\\
	&=\sqrt{{\rm Tr}(\hat{P}_{k-1}A^{\rm T}A)}+\sqrt{{\rm Tr}(\bar{P}_k)}+\sqrt{{\rm Tr}(Q)}\\
	&\leq\|A\|\sqrt{{\rm Tr}\hat{P}_{k-1}}+\sqrt{{\rm Tr}(\bar{P}_k)}+\sqrt{{\rm Tr}(Q)}.
\end{align*}

Matrix $\bar{P}_k$ has $2^n$ possible values, which is determined by the event-triggering conditions ${\bm \gamma}$ according to the definition of $\bar{P}_k$ in equation \eqref{eq: OQK} and the definition of $Q_k(a)$ in equation \eqref{eq: Qka}, where ${\bm \gamma}$ is a series of event-triggering conditions involved by $Q_k(a)$.

For the case that $\|A\|<1$, we have
\begin{equation*}
	\lim\limits_{k\rightarrow+\infty}\sqrt{{\rm Tr}(\hat{P}_k)}\leq\frac{\max\limits_{\bm \gamma}\sqrt{{\rm Tr}(\bar{P})}+\sqrt{{\rm Tr}(Q)}}{1-\|A\|},
\end{equation*}
which completes the proof.\hfill\QEDopen
\end{pf}
%
\begin{rem}
	Theorem \ref{thm: trP} states the asymptotic property of trace of the posteriori ellipsoidal estimation ${\bm x}_k\in\hat{X}_{k}=\mathcal{E}(\hat{\bm x}_k,\hat{P}_k)$. The condition $\|A\|<1$ leads to the asymptotic boundedness of ${\rm Tr}(\hat{P}_k)$, and thus the boundedness of estimation error between ${\bm x}_k$ and $\hat{\bm x}_k$, since ${\bm x}_k$ belongs to a bounded ellipsoid with center $\hat{\bm x}_k$ and the size matrix $\hat{P}_k$.
	\end{rem}

\begin{rem}\label{rem:extend}
	In this work, the $\epsilon$-observability and set-membership state estimator are discussed for systems with one dimensional measurement $y_k\in\mathbb{R}$ in Theorem \ref{thm: epob} and Theorem \ref{thm: trP}, respectively. The measurement of the system is designed to be one dimensional to make it convenience to establish the relationship between the column full rank and invertible of matrix $O$. For a system with multiple outputs, the measurement matrix $C$ satisfies $C\in\mathbb{R}^{m\times n}$, and thus the observability matrix satisfies $O=[C^{\rm T},(CA)^{\rm T},\ldots,(CA^{n-1})^{\rm T}]^{\rm T}\in\mathbb{R}^{mn\times n}$. For column full rank matrix $O\in\mathbb{R}^{mn\times n}$, an invertible square matrix can be obtained by choosing appropriate rows of matrix $O$. Then the proposed $\epsilon$-observability criterion and the designed set-membership state estimator can be extended to systems with $m$ dimensional measurements $y_k\in\mathbb{R}^m$.
\end{rem}

\section{Numerical Example}\label{sec:numexam}
In this section, numerical experiments are performed to verify the set-membership state estimator proposed in Algorithm \ref{alg: filter} and the asymptotic boundedness property proposed in Theorem \ref{thm: trP}. For convenience of illustrating the ellipsoidal sets utilized in this work, a second-order system with the form of system \eqref{eq: sys with w} is chosen with parameters:
\begin{align*}
	A&=\left[\begin{array}{cc}
		0.75&0.2\\
		0.5&0.3
	\end{array}\right],~~
	C=\left[0.5~~ 0.5\right],\\
	Q&=\left[\begin{array}{cc}
		5&0\\
		0&5
	\end{array}\right],~~~~R=0.5.
\end{align*}

The numerical experiment is performed with a deterministic send-on-delta mechanism as
\begin{equation*}
	\gamma_k=\left\{\begin{array}{ll}
		1, &{\text {if}}~ (y_k-y_{\tau k})^2>\Gamma,\\
		0, &{\text {if}} ~(y_k-y_{\tau k})^2\leq\Gamma,
	\end{array}\right.
\end{equation*}
where $\Gamma=0.6$ is the triggering threshold, $y_{\tau k}$ is the previously transmitted measurement. The set-valued measurement information inferred based on the received measurement and event-triggered conditions is 
\begin{equation*}
	\mathcal{Y}_{ok}=\left\{\begin{array}{ll}
		\left\{y:(y-y_{\tau k})^{\rm T}\Gamma_e^{-1}(y-y_{\tau k})\leq1\right\}	,& \gamma_k=1,\\
		\left\{y:(y-y_{\tau k})^{\rm T}\Gamma^{-1}(y-y_{\tau k})\leq1\right\},	& \gamma_k=0,
		\end{array}\right.
\end{equation*}
where $\Gamma_e=0.0001$.

Two performance indexes are defined to evaluate the performance of the proposed state estimator as estimation performance and communication rate. Let $E_d$ be the estimation error and $\eta$ be the communication rate defined as
\begin{equation*}
	E_d=\frac{1}{N}\sum\limits_{k=1}^N\|{\bm x}_k-\hat{\bm x}_k\|_2,~~\eta=\frac{1}{N}\sum\limits_{k=1}^N\gamma_k,
\end{equation*}
where $\|\cdot\|_2$ is the euclidean norm.

\begin{figure}[htbp]
	\centering
	\includegraphics[width=\linewidth]{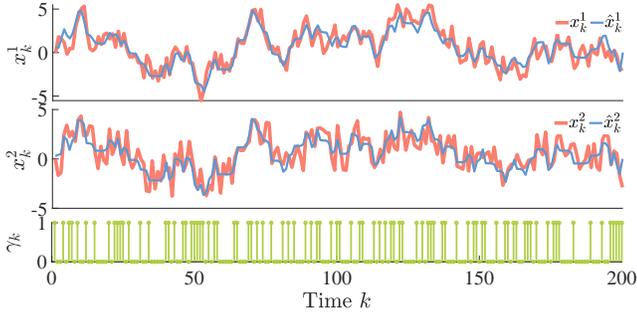}
	\caption{State estimation and event-triggering condition.}
	\label{fig:fig1}
\end{figure}

For the event-triggered state estimation problem with the parameters aforementioned, the simulation results and the ellipsoids introduced in this work are shown as follows. Fig. \ref{fig:fig1} shows the state estimation results and the event-triggering conditions, where the estimation error $E_d$ equals to $1.2261$ and the communication rate $\eta$ is $0.4472$. It can be seen in the first two subplots of Fig. \ref{fig:fig1} that the error of estimation $\hat{\bm x}_k$ is small even when there is no measurement information transmitted through the event-triggered communication channel, which verifies the estimation performance of the proposed stated estimator developed in Algorithm \ref{alg: filter} with the reduced communication rate. 

\begin{figure}[htbp]
	\centering
	\includegraphics[width=\linewidth]{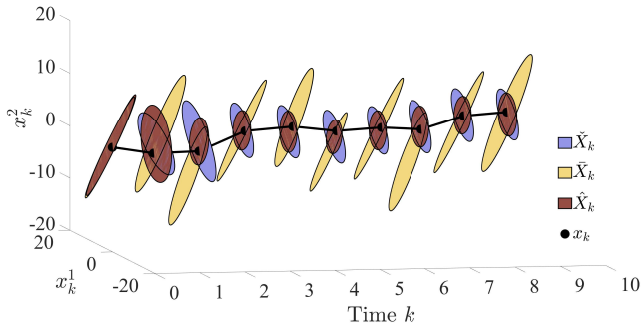}
	\caption{Measurements, prior, and posterior information ellipsoids in the process of state estimation.}
	\label{fig:fig2}
\end{figure}

During performing  Algorithm \ref{alg: filter}, three ellipsoids are obtained at every instant (except for the initial state ${\bm x}_0$, which are calculated according to equation \eqref{eq: x0est}), which are shown in Fig. \ref{fig:fig2}. To avoid complex and indistinct overlapping of the curves, only the ellipsoids for first $10$ states ${\bm x}_k, k\in\{0,1,\ldots,9\}$ are plotted. In Fig. \ref{fig:fig2}, the yellow ellipsoids $\bar{X}_k$ are the range of the state inferred according to the set-valued measurement information as equation \eqref{eq: xkhou}, and the \color{black} ellipsoids $\check{X}_k$ are the range of the state inferred according to the posterior range of the last state as equation \eqref{eq: xkqian}. The real state ${\bm x}_k$ belongs to the intersection of the ellipsoids $\bar{X}_k$ and $\check{X}_k$, which is shown as black points in Fig. \ref{fig:fig2}. The posterior ellipsoids $\hat{X}_k$ are obtained as external estimation of the intersection of ellipsoids $\check{X}_k$ and $\bar{X}_k$, which are plotted as red ellipsoids. The simulation results that the real states belong to the posterior estimated ellipsoids verify the validity of the designed estimator.

\begin{figure}[htbp]
	\centering
	\includegraphics[width=\linewidth]{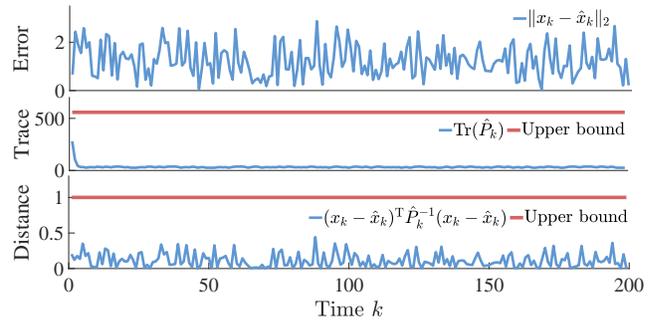}
	\caption{Verification of the performance of estimator and Theorem \ref{thm: trP}.}
	\label{fig:fig3}
\end{figure}

Moreover, to show the performance of the designed state estimator and the bound of the state estimation error, Fig. \ref{fig:fig3} is shown. In the first subplot of the Fig. \ref{fig:fig3}, the error of the state estimations $\|{\bm x}_k-\hat{\bm x}_k\|_2$ is plotted. The average estimation error equals $1.2261$, which verifies the effectiveness and accuracy of the proposed Algorithm \ref{alg: filter}. In the second subplot of the Fig. \ref{fig:fig3}, trace of the size matrix of posterior estimated ellipsoids is shown, and the asymptotic upper bound proposed in Theorem \ref{thm: trP} equals to 557.8243, which is plotted in the second subplot as a red curve. It can be observed from the second subplot that ${\rm Tr}(\hat{P}_k)$ is much smaller than the upper bound proposed in Theorem \ref{thm: trP}, which verifies the validity of the proposed upper bound.

In order to verify whether the true states of the systems ${\bm x}_k$ belong to the posterior ellipsoid $\hat{X}_k$ obtained using Algorithm \ref{alg: filter}, the generalized distance between ${\bm x}_k$ and $\hat{{\bm x}}_k$, $({\bm x}_k-\hat{\bm x}_k)^{\rm T}\hat{P}_k^{\rm T}({\bm x}_k-\hat{\bm x}_k)$, is shown in the third subplot of Fig.~\ref{fig:fig3} as a blue curve. It can be observed from the third subplot of Fig.~\ref{fig:fig3} that the generalized distance $({\bm x}_k-\hat{\bm x}_k)^{\rm T}\hat{P}_k^{\rm T}({\bm x}_k-\hat{\bm x}_k)$ is smaller than $1$, which is equivalent to ${\bm x}_k\in\hat{X}_k$ and confirms the effectiveness of the proposed method.

\section{Conclusion}\label{sec:conc}
In this work, we evaluate the observability of LTI systems with event-triggered communication channels utilizing a set-membership approach. A new notion, $\epsilon$-observability, is proposed as a generalization of classical observability, which focuses on the potential possibility of determining a set containing all possible values of the initial state. To judge the proposed $\epsilon$-observability, a testing criterion and the corresponding calculation method for parameter $\epsilon$ are developed using the intersection and sum of ellipsoids, based on which an event-triggered set-membership state observer is designed. The asymptotic property of the designed observer is analyzed and verified by numerical examples. In our current study, we focus on  deterministic event-triggering mechanisms. In our next step, stochastic event-triggering conditions will be further investigated.


%


\bibliographystyle{elsarticle-harv}
\bibliography{draft20211214}


\end{document}